\documentclass[]{aa}
\usepackage{graphicx}
\usepackage{txfonts}
          
\begin{document}  
  
\title{The evolution of white dwarfs with a varying gravitational constant}
  
\author{L. G. Althaus\inst{1,2,3},
        A. H. C\'orsico\inst{1,2,3},
        S. Torres\inst{4,5},
        P. Lor\'en--Aguilar\inst{4,5},
        J. Isern\inst{6,5},
        \and 
        E. Garc\'{\i}a--Berro\inst{4,5}}

\offprints{E. Garc\'{\i}a--Berro}  
  
\institute{Facultad de Ciencias Astron\'omicas y Geof\'{\i}sicas,  
           Universidad  Nacional de La Plata,
           Paseo del  Bosque s/n,  
           (1900) La Plata, 
           Argentina\
           \and
           Instituto de Astrof\'{\i}sica La Plata, 
           IALP, CONICET-UNLP,
           Argentina\
           \and 
           Member of the Carrera del Investigador Cient\'{\i}fico y 
           Tecnol\'ogico, CONICET, 
           Argentina\
           \and
           Departament de F\'\i sica Aplicada, 
           Universitat Polit\`ecnica de Catalunya,  
           c/Esteve Terrades, 5,  
           08860 Castelldefels,  
           Spain\
           \and
           Institute for Space Studies of Catalonia,
           c/Gran Capit\`a 2--4, Edif. Nexus 104,   
           08034  Barcelona, 
	   Spain\
           \and
           Institut de Ci\`encies de l'Espai (CSIC),
           Campus UAB, 08193 Bellaterra, 
	   Spain\\
\email{althaus,acorsico@fcaglp.unlp.edu.ar; 
       santi,loren,garcia@fa.upc.edu; 
       isern@ieec.cat}}
    
\date{\today}  
  
\abstract{Within the theoretical  framework of some modern unification
          theories   the  constants   of  nature   are   functions  of
          cosmological  time.  White dwarfs  offer the  possibility of
          testing  a possible  variation of  $G$ and,  thus,  to place
          constraints to these theories.}
         {We present full white dwarf evolutionary calculations in the
          case that  $G$ decreases with time.}
         {White dwarf evolution is  computed in a self-consistent way,
          including  the  most  up-to-date physical  inputs,  non-gray
          model atmospheres  and a detailed  core chemical composition
          that results  from the calculation of the  full evolution of
          progenitor stars.}
         {We find that the mechanical structure and the energy balance
          of white dwarfs  are strongly modified by the  presence of a
          varying  $G$. In  particular, for  a rate  of change  of $G$
          larger than $\dot{G}/G$=$-1  \times 10^{-12}$ yr$^{-1}$, the
          evolution  of cool  white dwarfs  is markedly  affected. The
          impact of  a varying  $G$ is more  notorious in the  case of
          more massive white dwarfs.}
         {In view of the recent results reporting that a very accurate
          white  dwarf cooling  age can  be  derived for  the old  and
          metal-rich open  cluster NGC  6791, our study  suggests that
          this  cluster could be  a potential  target to  constrain or
          detect a hypothetical secular variation of $G$}

  
\authorrunning{L. G. Althaus et al.}  
  
\titlerunning{White dwarf evolution with a varying gravitational constant}  

\maketitle  
  
   
\section{Introduction}  
\label{intro}  

One  of  the principles  of  General  Relativity  --- the  equivalence
principle  ---  requires  that  the fundamental  constants  should  be
independent of location.  However, in several modern grand-unification
theories,  the constants  of nature  are supposed  to be  functions of
low-mass   dynamical   scalar    fields   ---   see,   for   instance,
Lor\'en--Aguilar  et al.   (2003)  and references  therein.  If  these
theories are correct,  we expect them to experience  slow changes over
cosmological  timescales, and  thus to  also depend  on  location.  In
recent years, several constraints have been placed on the variation of
the fine structure  constant --- see the review  papers of Uzan (2003)
and Garc\'\i a--Berro et al.   (2007) for recent revisions of the most
stringent upper  limits to  their rate of  change.  In  sharp contrast
with the vivid  debate about whether (or not) there  is evidence for a
varying  fine  structure  constant,  relatively few  works  have  been
devoted to study a hypothetical variation of $G$.  The reason probably
lies  on the  intrinsic  difficulty  of measuring  the  value of  this
constant  (Mohr  et al.   2008).   Actually,  $G$  is the  fundamental
constant for  which we have  the less accurate determination,  and the
several measures of $G$  differ considerably. The tightest constraints
on the rate  of variation of $G$ come from  Lunar Laser Ranging, $\dot
G/G\la (2\pm7) \times 10^{-13}$~yr$^{-1}$ (M\"uller \& Biskupek 2007),
but  these are  purely local.  At intermediate  cosmological  ages the
Hubble diagram of Type Ia supernovae can also be used to constrain the
rate  of  variation  of   the  gravitational  constant,  $\dot  G/G\la
10^{-11}$~yr$^{-1}$ at $z\sim 0.5$ (Gazta\~naga et al. 2002). Finally,
Big Bang  Nucleosynthesis also provides limits  on possible variations
in Newton's  Constant, $-3\times10^{-13}~{\rm yr}^{-1}  \la (\dot G/G)
\la 4 \times 10^{-13}~{\rm yr}^{-1}$ (Copi et al. 2004).

White  dwarfs   provide  an   independent  way  of   constraining  any
hypothetical variation  of $G$.  There  are several reasons  for this.
First, white dwarfs are extremely long-lived stars.  Thus, the effects
of a  varying $G$ can become  prominent, even for very  small rates of
change.  Second,  white dwarfs are the end-point  of stellar evolution
for the  vast majority  of stars.  Hence,  they are  numerous.  Third,
white  dwarfs  are  compact  objects,  and  their  structure  is  very
sensitive to the precise value of $G$. Finally, the evolution of white
dwarfs is relatively  well understood, and can be  well described as a
simple gravothermal process.  Hence, for sufficiently low temperatures
their luminosity is derived entirely  from a close balance between the
thermal  and the  gravitational energies.   Consequently,  a secularly
varying $G$  largely affects the gravothermal balance  of white dwarfs
and, thus, their luminosities.

\begin{table*}
\caption{White dwarf evolutionary sequences computed in this work.  We
         list the white  dwarf stellar mass (in solar  units) and, for
         each  value of  the rate  of change  of $G$,  $\dot{G}/G$ (in
         units  of  yr$^{-1}$),  the  initial  value  of  $G$  at  the
         beginning of  the white dwarf cooling  phase, $G_i/G_0$.  The
         numbers   in   brackets    give   the   stellar   luminosity,
         $\log(L/L_{\sun})$,  at  which   the  present  value  of  the
         gravitational constant, $G_0$, occurs.}
\begin{center}
\begin{tabular}{cccc}
\hline
\hline
\\
$M_{\rm WD}/M_{\sun}$ & \multicolumn{3}{c}{$G_i/G_0$}\\
\cline{2-4}
& 
$\dot{G}/G$=$-5 \times 10^{-11}$ &  
$\dot{G}/G$=$-1 \times 10^{-11}$ &  
$\dot{G}/G$=$-1 \times 10^{-12}$\\
\hline
0.525 &  1.40 ($-4.40$)  & 1.10 ($-4.37$)  &  1.010 ($-4.33$)  \\
0.525 &  1.30 ($-4.20$)  & 1.05 ($-4.05$)  &  1.005 ($-4.00$)  \\
0.525 &  1.20 ($-4.05$)  & 1.02 ($-3.66$)  &                   \\
0.525 &  1.10 ($-3.77$)  &                 &                   \\
\hline
0.609 &  1.50 ($-4.83$)  & 1.20 ($<-5$)    &  1.100 ($<-5$)    \\ 
0.609 &  1.40 ($-4.40$)  & 1.10 ($-4.30$)  &  1.050 ($<-5$)    \\
0.609 &  1.30 ($-4.19$)  & 1.05 ($-4.02$)  &  1.020 ($<-5$)    \\ 
0.609 &  1.20 ($-4.02$)  & 1.02 ($-3.60$)  &                   \\
0.609 &  1.10 ($-3.68$)  &                 &                   \\ 
\hline
1.000 &  1.24 ($-4.64$)  & 1.10 ($-4.55$)  &  1.020 ($<-5$)    \\ 
1.000 &  1.20 ($-4.12$)  & 1.05 ($-3.68$)  &  1.010 ($-4.30$)  \\
1.000 &  1.10 ($-3.23$)  & 1.02 ($-3.10$)  &  1.005 ($-3.56$)  \\ 
\hline
\hline
\end{tabular}
\end {center}
\label{secuencias}
\end{table*}  

One  of the ways  in which  white dwarfs  can be  used to  constrain a
variation  of $G$  considers the  dependence  of the  secular rate  of
change  of the period  of pulsation  of variable  white dwarfs  on its
cooling rate (Benvenuto et al.  2004; Biesiada \& Malec 2004).  It has
been shown  that their secular rate  of change of the  period not only
depends on  the cooling rate  but also on  the rate of change  of $G$.
Benvenuto  et al.   (2004) applied  this  method to  the well  studied
variable  white dwarf  G117-B15A and  obtained a  rather  loose bound,
$-2.5 \times  10^{-10}$ yr$^{-1}\le \dot{G}/G \le  0$.  Another method
to constrain  a possible variation  of $G$ is  to use the  white dwarf
luminosity  function.   The  number  counts  of  white  dwarfs  depend
sensitively on the characteristic cooling  time of white dwarfs in the
corresponding luminosity interval, so does the position of the cut-off
of the white dwarf  luminosity function at low luminosities.  Garc\'\i
a--Berro et  al.  (1995)  used a simplified  treatment to  check which
could  be the  effects of  a  slowly varying  $G$ in  the white  dwarf
luminosity function. They assumed that $\dot{G}/G$ was small enough to
ensure that white dwarfs have  time to adjust its mechanical structure
to the present  value of $G$ in a timescale much  shorter than that of
the  cooling timescale.   Under this  assumption,  energy conservation
leads to (Garc\'\i a--Berro et al. 1995):

\begin{equation}
L + L_{\nu}= - \frac{d B}{dt} + \frac{\dot{G}}{G} \Omega.
\label{lumiwd}
\end{equation}

\noindent  where  $B$  is  the  binding energy  of  the  white  dwarf,
$B=U+\Omega$, $U$ is the total  internal energy, $\Omega$ is the total
gravitational  energy and  $L_{\nu}$ is  the neutrino  luminosity. The
evolution of the luminosity was  then obtained from a series of static
models,   not  from  fully   evolutionary  calculations,   assuming  a
relationship between the luminosity  and the core temperature obtained
from fits to evolutionary  calculations with constant $G$.  Despite of
this  simplified treatment  they were  able to  obtain an  upper limit
$\dot{G}/G\le   -(1.0\pm  1.0)\times   10^{-11}$   yr$^{-1}$.   Later,
Biesiada  \&  Malec  (2004)  casted  doubts  on  this  treatment.   In
particular,  they claimed  that the  last term  in Eq.~(\ref{lumiwd}),
involving  $\dot G/G$,  cancels  when the  expansion  work is  further
expanded.  However,  their criticism was  not correct because  all the
relevant  terms in  the  energy conservation  equation were  carefully
taken into account in Garc\'\i a--Berro et al. (1995).

What neither  Garc\'\i a--Berro  et al. (1995)  nor Biesiada  \& Malec
(2002) took into account is the fact that the relationship between the
core temperature  and the luminosity  of white dwarfs also  depends on
$G$. For instance, adopting  the simplified Mestel cooling law (Mestel
1952) it turns out that $L\propto  G$.  This means that a full stellar
evolutionary code is  needed if an accurate treatement  of white dwarf
cooling when $G$  changes with time is to be  done.  To our knowledge,
the only calculations of cooling  white dwarfs with a time-varying $G$
employing  an  up-to-date  stellar  evolutionary  code  are  those  of
Benvenuto et al.  (1999), but  their analysis turned out to be flawed.
Specifically,  in this  work the  energy conservation  equation  for a
running  $G$ included  the differential  version of  the last  term of
Eq.~(\ref{lumiwd}), which  is not  correct.  In summary,  although the
evolutionary properties of white dwarfs can be potentially used to set
upper bounds  to the rate of  variation of $G$, very  few studies have
been done  up to now, and those  already done are not  complete or are
flawed. Moreover,  none of these works  took into account  that if $G$
varies its value  in the past differs from its  present value, and all
the  calculations were  done  using  the present  value  of $G$,  thus
neglecting a potentially important  effect. This paper aims at filling
this gap.

\section{Input physics}

In this work we follow in a self-consistent way the evolution of white
dwarfs in the case of a varying $G$.  In the interest of simplicity we
have  assumed that $\dot{G}/G$  remains constant  with time.   All the
calculations   have  been   done  using   the  {\tt   LPCODE}  stellar
evolutionary code  --- see  Althaus et al.   (2010) and Renedo  et al.
(2010) for recent applications of this code.  The main physical inputs
considered in  {\tt LPCODE} comprise the following.   We have included
$^{22}$Ne  diffusion  and  its   associated  energy  release  ---  see
Garc\'{\i}a--Berro et al.  (2008; 2010) and Althaus et al.  (2010) for
details.  The energy sources arising from crystallization of the white
dwarf core --- namely, the release of latent heat and of gravitational
energy  associated  with   carbon-oxygen  phase  separation  (Garc\'\i
a--Berro et  al. 1988; Isern  et al. 1997;  2000) --- are  fully taken
into account  following the treatment  of Althaus et al.   (2010).  We
used the carbon-oxygen phase  diagram of Segretain \& Chabrier (1993).
Other phase diagrams for  the carbon-oxygen mixture have been recently
computed (Horowitz et al. 2010), but  as long as we are concerned with
relative differences in  the white dwarf cooling times  when a running
$G$ is adopted, the impact of  a different choice of the phase diagram
on   the  bounds   on   $\dot   G/G$  should   be   minor.   For   the
$^{12}$C$(\alpha,\gamma)^{16}$O    reaction   rate   we    adopt   the
prescription of Angulo  et al. (1999). Again, we  stress that although
the  precise value of  this poorly  known reaction  rate ---  see, for
instance Assun{\c c}{\~a}o  et al.  (2006) --- is  important to obtain
the  chemical stratification of  white dwarf  progenitors and  thus to
compute absolute  white dwarf cooling  ages, the relative effect  of a
varying $G$ in  the models does not appreciably depend  on it.  We use
non-gray  model   atmospheres  to  provide   accurate  outer  boundary
conditions for our models.   Our atmospheres include non-ideal effects
in the  gas equation  of state and  chemical equilibrium based  on the
occupation     probability    formalism.     They     consider    also
collision-induced  absorption due to  H$_2$-H$_2$, H$_2$-He,  and H-He
pairs,  and the  Ly$\alpha$ quasi-molecular  opacity that  result from
perturbations of hydrogen atoms  by interactions with other particles,
mainly  H and  H$_2$  (Rohrmann  et al.   2010).   Our sequences  also
incorporate  element diffusion  in the  outer layers  (Althaus  et al.
2003).

The radiative opacities  used are those of the  OPAL project (Iglesias
\&  Rogers 1996),  and include  carbon- and  oxygen-rich compositions.
These  opacities  are  supplemented   at  low  temperatures  with  the
molecular opacities for pure hydrogen composition of Marigo \& Aringer
(2009). The  conductive opacities of  Cassisi et al. (2007)  have also
been   used.    Neutrino  emission   rates   for   pair,  photo,   and
bremsstrahlung processes are  those of Itoh et al.   (1996), while for
plasma processes we include the  treatment of Haft et al.  (1994).  We
use  the  equation  of state  of  Segretain  et  al.  (1994)  for  the
high-density regime.   For the low-density regime, we  used an updated
version of the equation of state of Magni \& Mazzitelli (1979). 

Given  that our  aim is  to compute  self-consistently the  cooling of
white dwarfs  in the  presence of  a varying $G$,  we write  the local
luminosity equation as

\begin{equation}
\frac{\partial L_r}{\partial m}=  -\epsilon_\nu - \frac{\partial u} 
{\partial t} + \frac{P}{\varrho^2}\, \frac{\partial \varrho}{\partial t}. 
\label{lumistandard}
\end{equation}

\noindent and we  allow $G$ to vary.  This is a  fair approach and has
been  adopted in  previous  studies of  this  kind (degl'Innocenti  et
al. 1996).  Consequently,  the density profile of the  white dwarf now
varies not  only because the white  dwarf cools, but  also because $G$
varies.  Naturally, this influences the white dwarf cooling ages.

\begin{figure}
\begin{center}
\includegraphics[clip,width=0.9\columnwidth]{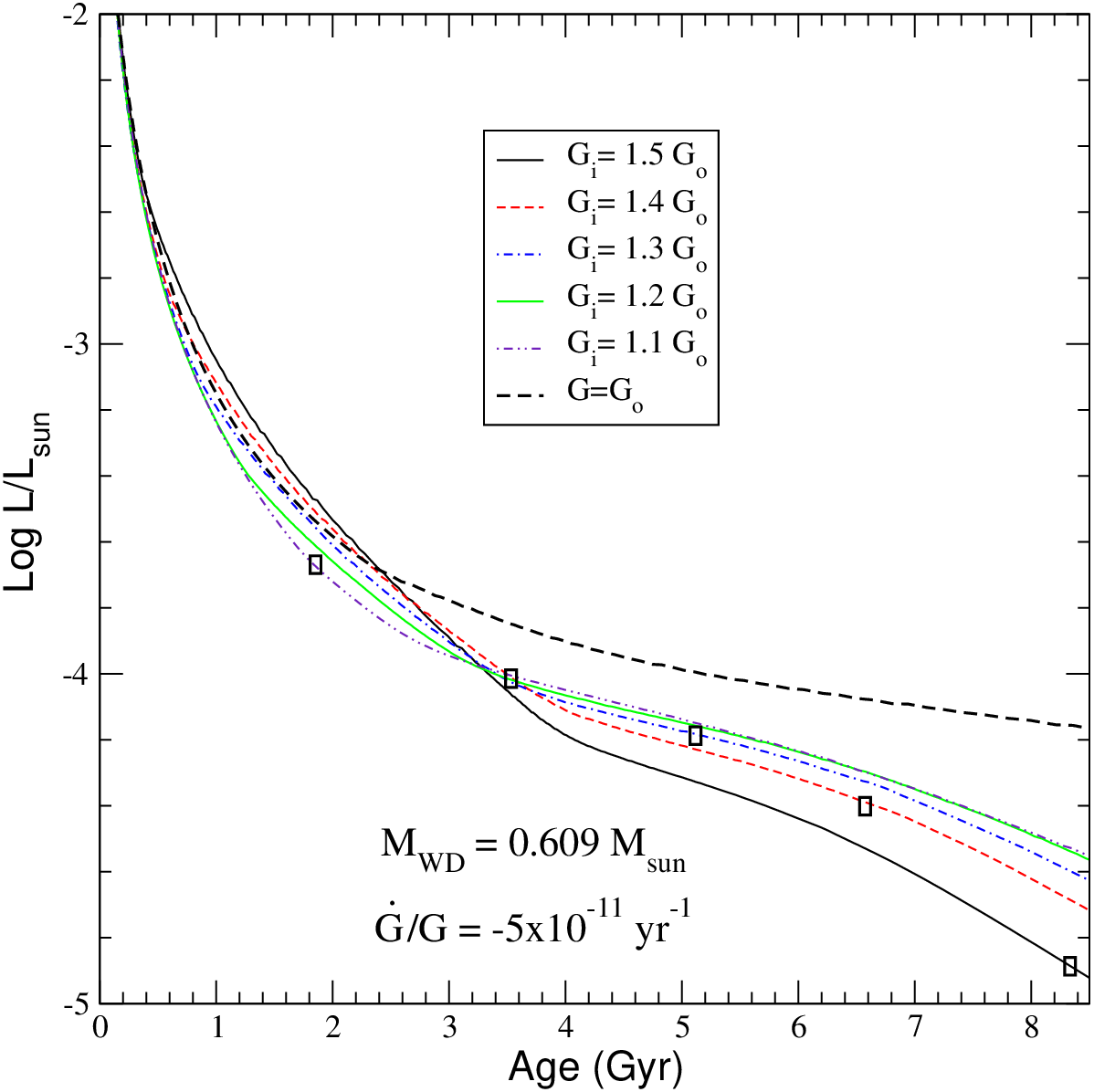}
\caption{Surface luminosity versus age  for a $0.609\, M_{\sun}$ white
         dwarf  assuming  $\dot{G}/G= -5\times10^{-11}$~yr$^{-1}$  and
         different initial values  of $G$ at the start  of the cooling
         track.  We also show  using square symbols the luminosity and
         age  at which  the present  value of  $G$ is  reached  in the
         cooling track.}
\label{evol06}
\end{center}
\end{figure}

\section{Evolutionary results}

We  have computed  the evolution  of white  dwarf sequences  of masses
0.525,  0.609 and $1.0\,  M_{\sun}$ considering  three values  for the
rate    of    change    of    $G$,   namely    $\dot{G}/G=-5    \times
10^{-11}$~yr$^{-1}$,  $\dot{G}/G=-1  \times  10^{-11}$~yr$^{-1}$,  and
$\dot{G}/G=-1  \times 10^{-12}$~yr$^{-1}$.   As it  will  become clear
later,  the  evolution  in the  case  of  a  varying $G$  is  strongly
dependent on the initial value of  $G$ at the beginning of the cooling
phase.  Accordingly,  for each value  of $\dot{G}/G$ we  have computed
several  sequences with  different  values of  $G_i/G_0$, where  $G_i$
stands for the  value of $G$ at the beginning of  the cooling track at
high effective temperature, and $G_0$ corresponds to the present value
of  $G$.   To obtain  starting  white  dwarf  configurations with  the
appropriate  value of $G$,  we simply  changed $G_0$  by $G_i$  at the
beginning  of  the  cooling  phase  and we  calculated  the  resulting
structure.   After  a  brief  transitory  stage, we  get  the  initial
configurations for our sequences.  This artificial procedure to obtain
the  initial  white  dwarf   configurations  is  justified  since  the
subsequent  evolution does  not depend  on the  way the  initial white
dwarf structures are obtained.  All  in all, we have computed 30 white
dwarf     evolutionary    sequences,     which    are     listed    in
Table~\ref{secuencias}. In  addition, in this  table we also  list the
surface luminosity (in  solar units) for which $G=G_0$.  Note that the
election of the values  of $G_i$ has been made in such  a way that the
present  value  of  $G$  occurs  at advanced  stages  of  white  dwarf
evolution, when the  surface luminosity ranges from $\log(L/L_{\sun})=
-3$  to $-5$,  the typical  luminosities of  field white  dwarfs.  All
these sequences  have been  computed down to  $\log(L/L_{\sun})=-5$, a
luminosity smaller than that of the dimmest field white dwarfs.

\begin{figure}
\begin{center}
\includegraphics[clip,width=0.9\columnwidth]{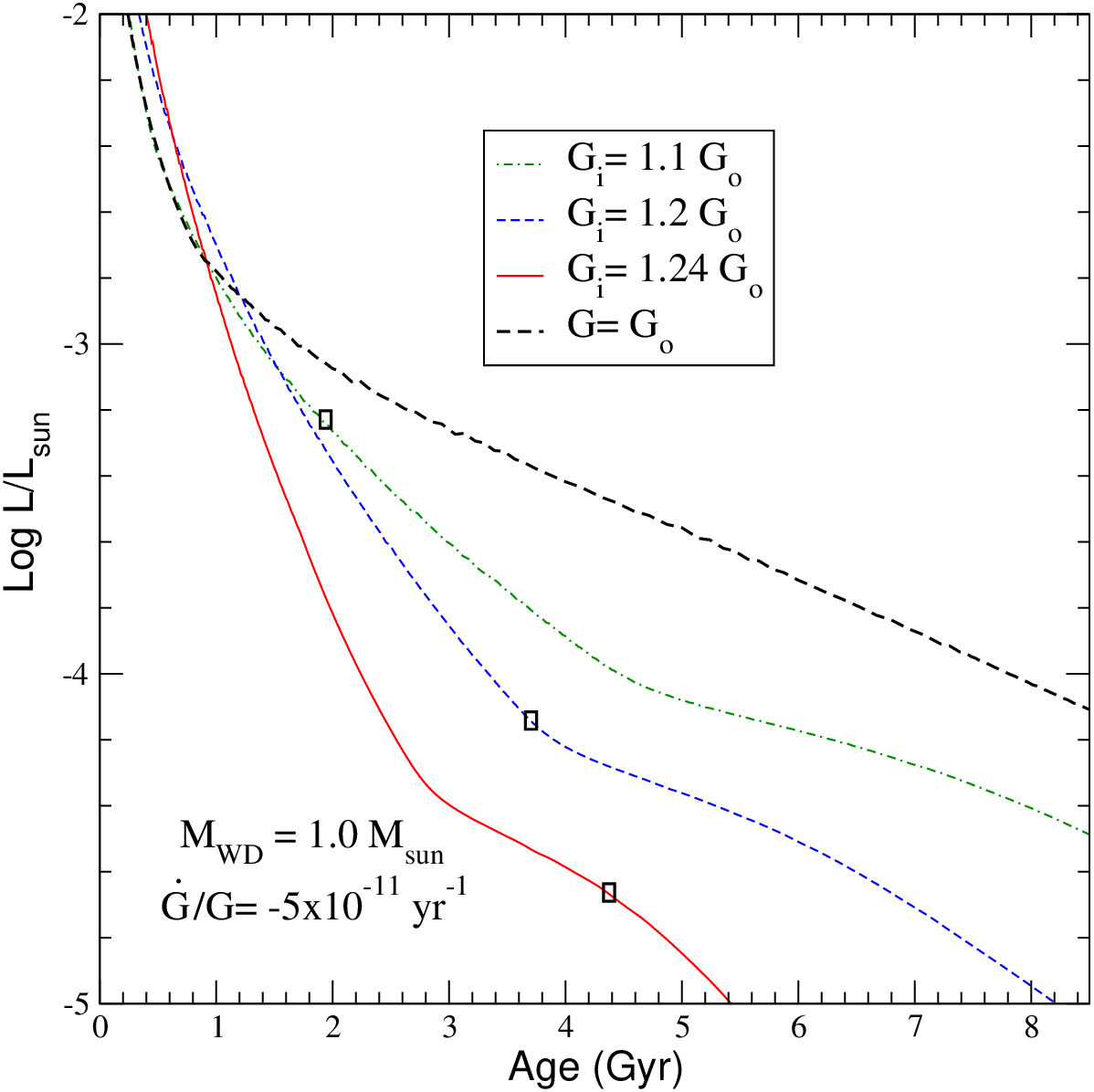}
\caption{Surface luminosity  versus age  for a $1.0\,  M_{\sun}$ white
         dwarf  assuming  $\dot{G}/G= -5\times10^{-11}$~yr$^{-1}$  and
         different initial values  of $G$ at the start  of the cooling
         track.  We also show  using square symbols the luminosity and
         age  at which  the present  value of  $G$ is  reached  in the
         cooling track.}
\label{evol10}
\end{center}
\end{figure}

Despite  of  the  small rates  of  change  of  $G$ adopted  here,  the
evolution of our white dwarf sequences is strongly modified.  Since we
have  adopted  $\dot{G}/G<0$,  the  central  density  of  the  cooling
sequences decreases with  time. This is at odds  with the evolution of
standard models, in which  white dwarfs cool at approximately constant
density.   Since the  energy of  cool white  dwarfs is  essentially of
gravothermal origin, this has important implications for the evolution
of old white dwarfs, and even  a small change in $G$ alters the energy
balance of the star and  thus its luminosity.  In addition, because of
the  larger  value of  $G$  at the  beginning  of  the cooling  track,
$^{22}$Ne sedimentation in the liquid phase and crystallization of the
white   dwarf  core   occur  at   larger  luminosities.    Hence,  the
gravitational  energy released  by  these processes  takes place  much
earlier, thus  strongly modifying the cooling rate  when compared with
that obtained in the case in which $G$ is constant.

The white dwarf cooling times are sensitive to the value of $G$ at the
beginning  of the  cooling track.   This can  be better  understood by
examining Figs.  \ref{evol06} and \ref{evol10} which show the relation
between the surface luminosity and  the age for the $0.609\, M_{\sun}$
and    $1.0\,    M_{\sun}$     white    dwarf    sequences    assuming
$\dot{G}/G=-5\times   10^{-11}$~yr$^{-1}$  and  different   values  of
$G_i/G_0$.  In  these figures  we also show  using square  symbols the
luminosity and age at which the present value of $G$ is reached in the
cooling track. Note that the  larger the value of $G_i/G_0$, the lower
the luminosity at which  $G_i=G_0$ occurs.  Obviously, this means that
if  the correct value  of $G$  is to  be obtained  at present  time an
initially  more compact  white dwarf  is  needed.  Also,  it is  worth
noting that  the cooling ages  in the case  of a varying $G$  are much
smaller at low luminosities than the cooling ages obtained when $G$ is
constant. Finally,  because of the  stronger gravity of  massive white
dwarfs, the impact  of a decreasing $G$ on  the cooling time increases
markedly with the stellar mass.

\begin{figure}
\begin{center}
\includegraphics[clip,width=0.9\columnwidth]{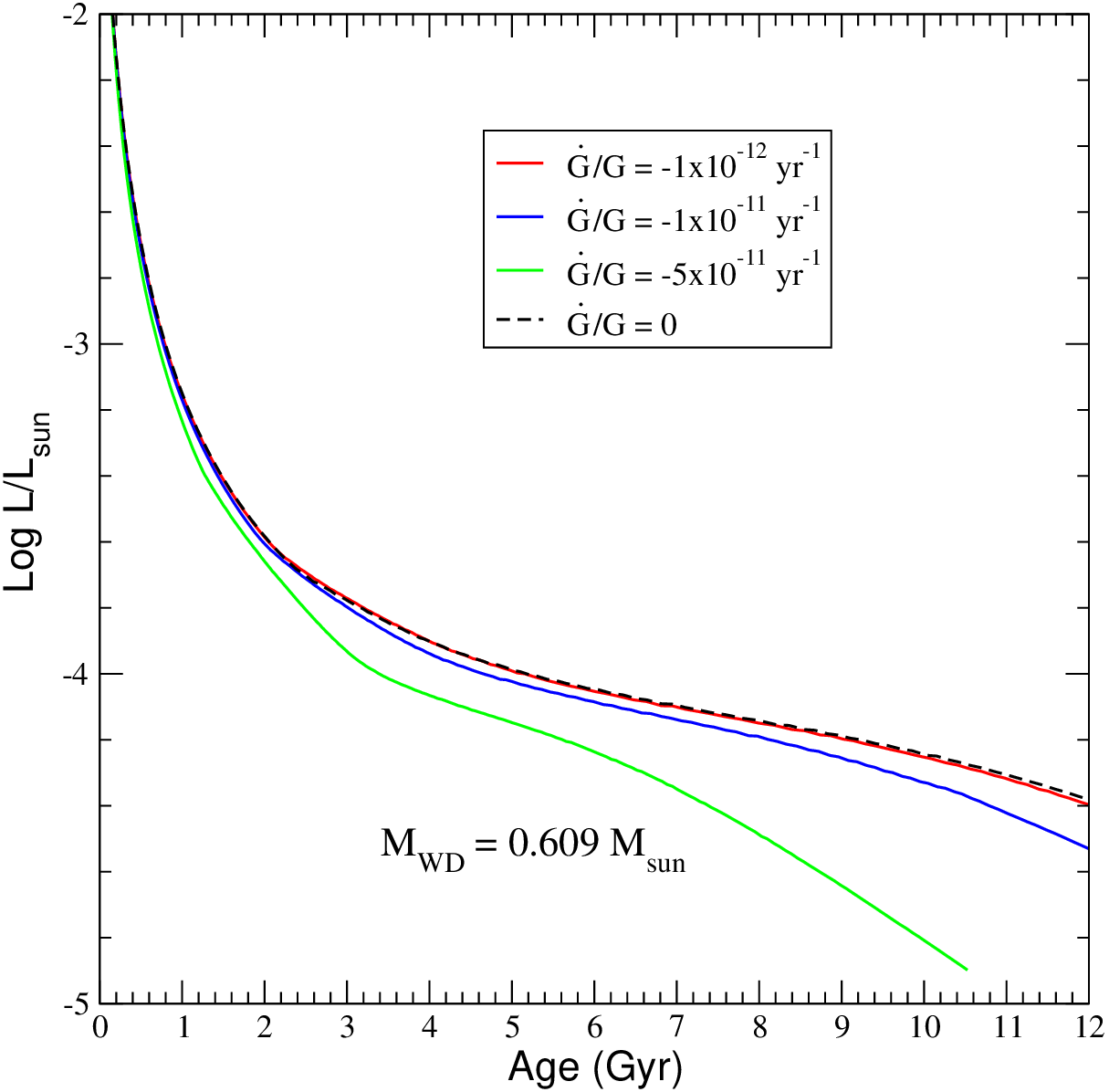}
\caption{Surface luminosity versus  age for several $0.609\, M_{\sun}$
         white   dwarf  sequences,   adopting   different  values   of
         $\dot{G}/G$. See text for details.}
\label{tresg06}
\end{center}
\end{figure}

\begin{figure}
\begin{center}
\includegraphics[clip,width=0.9\columnwidth]{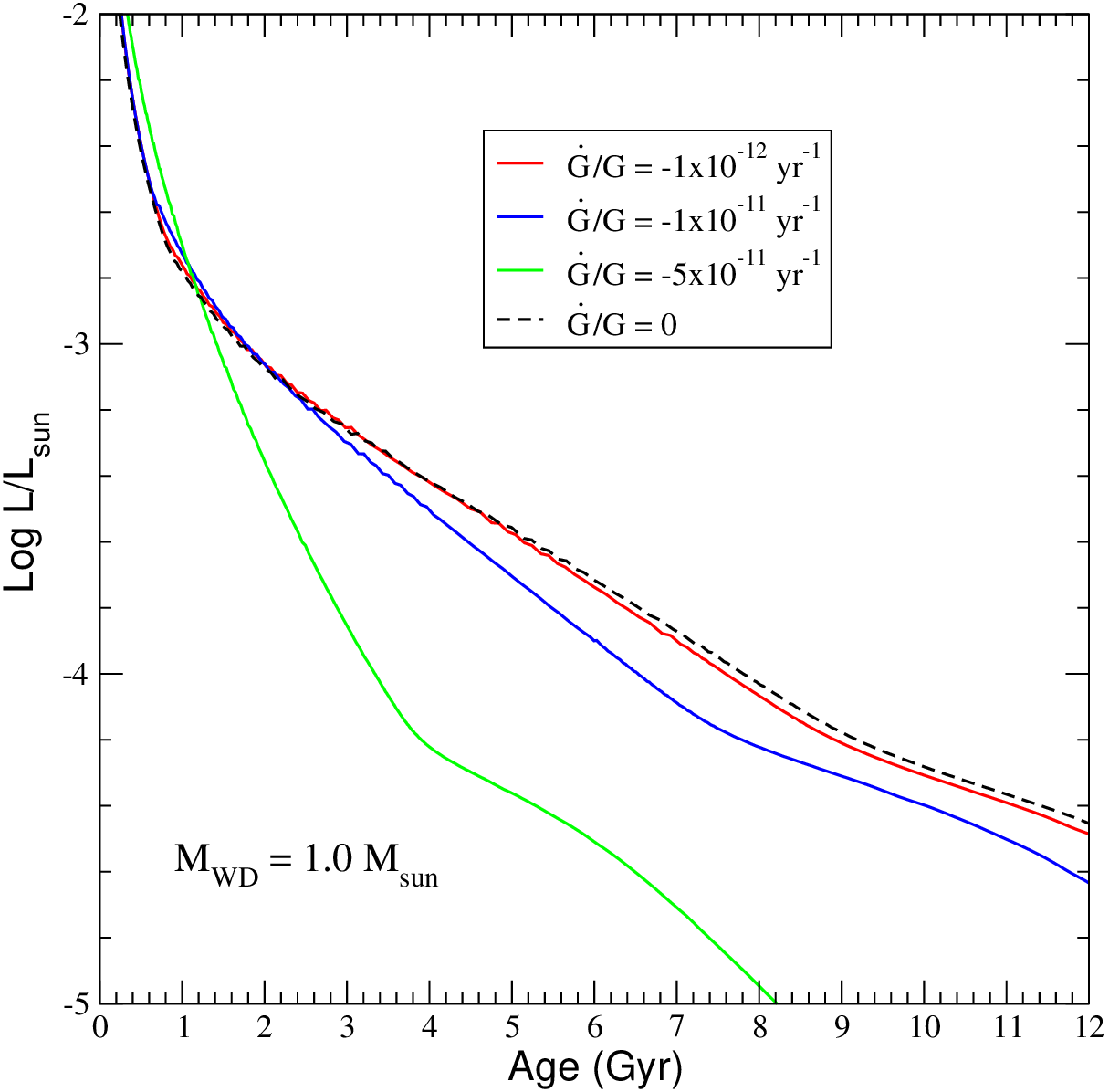}
\caption{Surface  luminosity versus age  for several  $1.0\, M_{\sun}$
         white   dwarf  sequences,   adopting   different  values   of
         $\dot{G}/G$. See text for details.}
\label{tresg1}
\end{center}
\end{figure}

To illustrate the dependence of cooling  on the rate of change of $G$,
we  show  in  Figs.    \ref{tresg06}  and  \ref{tresg1}  the  temporal
evolution  of the  luminosity for  the $0.609\,  M_{\sun}$  and $1.0\,
M_{\sun}$         white        dwarf         sequences        assuming
$\dot{G}/G=-1\times10^{-12}$~yr$^{-1}$,   $-1\times10^{-11}$~yr$^{-1}$,
and $-5\times10^{-11}$~yr$^{-1}$.  The initial values of $G$ have been
selected in such a way that for each sequence the present value of $G$
occurs  at $\log(L/L_{\sun})\approx  -4.0$.  To  further  quantify the
effects  of a varying  $G$ we  also show  in table~\ref{G0}  the white
dwarf cooling  ages at $\log(L/L_{\sun})= -4.0$  of these evolutionary
sequences and  the corresponding  ages for the  models in  which $\dot
G/G=0$. As can  be seen in these figures  and in table~\ref{G0}, there
is a marked dependence of the  white dwarf ages on the assumed rate of
change  of $G$.   This dependence  is more  notorious in  the  case of
massive   white    dwarfs.    However,   for    $\dot{G}/G=-1   \times
10^{-12}$~yr$^{-1}$  the evolution  is  almost indistinguishable  from
that  of the  standard  case  of constant  $G$,  particularly for  the
low-mass white  dwarf.  This value of $\dot{G}/G$  constitutes a lower
limit for the rate of change of $G$ above which we can expect that the
evolution of white dwarfs will be influenced by a varying $G$.

\section{Discussion and conclusions}

Recently, Garc\'\i  a--Berro et al. (2010) have  demonstrated that the
slow down  of the  white dwarf  cooling rate owing  to the  release of
gravitational  energy from  $^{22}$Ne sedimentation  and carbon-oxygen
phase separation upon crystallization  is of fundamental importance to
reconcile the age discrepancy in the very old, metal-rich open cluster
NGC  6791.  This  raises  the  possibility of  using  this cluster  to
constrain a hypothetical variation of $G$ with time. In a step forward
to do  this, in this work  we have computed  a new set of  white dwarf
evolutionary sequences in which we allow $G$ to vary.  We have assumed
that $\dot{G}/G$ remains  constant with time and we  have followed the
evolution of  white dwarf model  sequences of masses 0.525,  0.609 and
$1.0\, M_{\sun}$  considering three values  for the rate of  change of
$G$,  namely $\dot{G}/G=-5  \times  10^{-11}$~yr$^{-1}$, $\dot{G}/G=-1
\times      10^{-11}$~yr$^{-1}$,     and      $\dot{G}/G=-1     \times
10^{-12}$~yr$^{-1}$. To the  best of our knowledge these  are the only
self-consistent  evolutionary   sequences  of  cooling   white  dwarfs
incorporating the effects of a varying $G$.

\begin{table}
\caption{Cooling   times,  in   Gyr,  at   $\log(L/_{\sun})=-4.0$  for
         different values  of $\dot G/G$ and  two representative white
         dwarf masses.  For these  cooling sequences the present value
         of $G$ occurs at this value of the stellar luminosity.}
\begin{center}
\begin{tabular}{ccc}
\hline
\hline
\\
{$\dot{G}/G$}  & \multicolumn{2}{c} {$M_{\rm WD}/M_{\sun}$} \\
\cline{2-3}
&
$0.609$ &
$1.0$ \\
\hline
0&  5.200  & 7.798    \\
$-1 \times 10^{-12}$ &  5.148  & 7.502   \\
$-1 \times 10^{-11}$ &  4.677  & 6.718    \\
$-5 \times 10^{-11}$ &  3.435  & 3.311    \\
\hline
\hline
\end{tabular}
\end {center}
\label{G0}
\end{table}

We have found that the  mechanical structure and the energy balance of
cool white dwarfs  are strongly modified when a  slowly varying $G$ is
adopted,  and that  the  white  dwarf evolution  is  sensitive to  the
initial value of  $G$ at the beginning of  the cooling phase.  Because
of  the  compact nature  of  massive white  dwarfs,  the  impact of  a
decreasing $G$ on the cooling time increases markedly with the stellar
mass.    We    have   found   as    well   that   for   a    rate   of
$\dot{G}/G=-5\times10^{-11}$ yr$^{-1}$, the  cooling ages of dim white
dwarfs result much smaller than the cooling ages given by the standard
evolution at constant $G$.   The calculations presented here show that
the  white dwarf evolution  is very  sensitive to  the exact  value of
$\dot{G}/G$  and that  a detailed  fit to  the white  dwarf luminosity
function of NGC 6791 may place tight constraints on a possible rate of
change of  $G$. Last  but not least,  we mention that  calculations to
study the evolution of the progenitors of white dwarfs, which may also
affect the cooling times, are currently under way.

\begin{acknowledgements}
This   research   was   supported    by   AGAUR,   by   MCINN   grants
AYA2008--04211--C02--01  and  AYA08-1839/ESP,  by  the  ESF  EUROCORES
Program  EuroGENESIS  (MICINN grant  EUI2009-04170),  by the  European
Union   FEDER   funds,  by   AGENCIA:   Programa  de   Modernizaci\'on
Tecnol\'ogica BID 1728/OC-AR, and  by PIP 2008-00940 from CONICET. LGA
also  acknowledges a  PIV grant  of the  AGAUR of  the  Generalitat de
Catalunya.
\end{acknowledgements}

\end{document}